\documentclass[12pt]{iopart}
\usepackage{graphicx}
\begin{document}

\title{High $p_{T}$ Hadron Spectra at High Rapidity }

\author{Ramiro Debbe\dag\ for the BRAHMS Collaboration  }

\address{\dag\ Brookhaven National Laboratory, Upton NY, 11973 }

\begin{abstract}
We report the measurement of charged hadron production at different pseudo-rapidity values in deuteron+gold as well as
proton-proton collisions at $\sqrt{s_{NN}}=200GeV$ at RHIC. The nuclear modification factors $R_{dAu}$ and $R_{cp}$ are used 
to investigate new behaviors in the deuteron+gold system as function of rapidity and the centrality of the collisions respectively.

\end{abstract}




\section{Introduction}

The work reported in this presentation was motivated by the possibility of observing the onset of 
gluon saturation in
d+Au collisions at RHIC. Saturation of the parton densities as their fractional longitudinal momenta $\it x$
tends toward zero is widely accepted on the basis of hard limits set by unitarity 
(Froissart bound) \cite{Froissart}, but the proper scale for that
new regime, as well as its properties are the subject of intense study. Deep inelastic scattering of leptons 
on hadronic systems 
performed at
HERA and FNAL have shown that for small-$x$ values gluons are the dominant component of the hadron wave function 
 and that their densities grow as powers of $\frac{1}{x}$ \cite{HERAdata}.
Limits to that growth have been proposed in a phenomenological description of the cross section of the virtual 
photon 
exchanged in those reactions. A functional form of that cross section is proposed such that it  grows to 
become a constant beyond a scale that depends on $\it x$ as $Q_{s}^{2} \sim \frac{1}{x^{\lambda}} $ where 
$\lambda = 0.2 - 0.3$ is the only free parameter with values extracted from fits to 
data \cite{Golec,Lambda}.
A QCD based theory of dense partonic systems accessible with the new  high energy colliders has 
been developed \cite{McLerranVenu}. And the new regime it strives to describe is called the
 Color Glass Condensate (CGC). This theory offers a novel way of calculating nuclear properties at high energies based 
on classical field theory techniques.

The formalism of saturation can also describe p+A systems and the characteristic scale in transverse momentum 
 is directly proportional to the density
of partons in the ion: $Q_s^2 \sim \frac{A^{\frac{1}{3}}} {x^{\lambda}}$ and recalling the relation between the 
rapidity of a parton that carries the fraction x of longitudinal momentum and the rapidity of the hadron: 
$y_{parton} = y_{hadron} - ln(\frac{1}{x})$
one can write  $Q_s^2 \sim A^{\frac{1}{3}} e^{\lambda y}$. RHIC offers the advantage of combining 
high energy and high A 
ions (197 for gold) and
BRAHMS, one of the four experiments currently collecting data at RHIC, can study particle production at 
 high rapidity.

\section{Experimental setup}

The data presented in this report were collected with both BRAHMS spectrometers, the mid-rapidity 
spectrometer (MRS) and the 
front section of the forward spectrometer (FFS), complemented with an event characterization  system used to determine the 
geometry of the collisions. 
A detailed description of the BRAHMS experimental setup can be found in \cite{BRAHMSNIM}. 
However, the low multiplicity
of charged particles in the proton or light ion reactions required an extension of the basic apparatus with a set of
scintillator counters (called INEL detectors).  These detectors \cite{BRAHMS_daumult} cover pseudo-rapidities in the
range: $3.1 \leq \mid \eta \mid \leq 5.29$, and  
define a minimum biased trigger.
This trigger is estimated to select $\approx 91\% \pm 3\%$ of the 2.4 barns d+Au 
inelastic cross section and $71\% \pm 5\%$ of the total inelastic proton-proton 
cross section of 41 mb. 
The INEL detector  
was also used to select events with collision 
vertex within $\pm 15$ cm of the nominal collision point with a resolution of 5 cm.

The centrality of the collision was extracted from the multiplicity of the event measured within the
angular region $\mid \eta \mid \leq 2.2$ with a combination of silicon and scintillator counters \cite{BRAHMS_Tiles}.
The results reported here were obtained without making use of the full complement of particle identification
detectors. Ongoing analysis of the data includes particle identification and the use of the complete forward
spectrometer FS.

\section{Spectra}

\begin{figure}[!ht]
\begin{center}
\resizebox{0.8\textwidth}{!}
           {\includegraphics{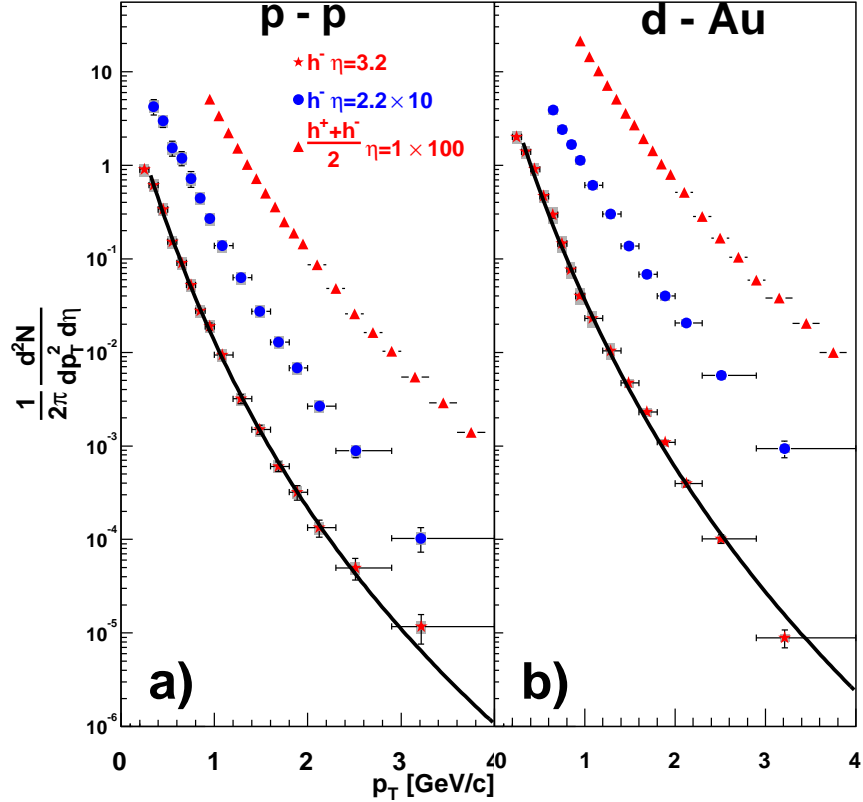}}
\end{center}
\caption{\label{fig:spectra}Spectra for charged hadrons at different 
pseudo-rapidities. Panel a shows the spectra obtained from proton-proton collisions and panel b those from d+Au collisions. 
The top  most distributions in both panels correspond to the invariant yields of $\frac{h^+ + h^-}{2}$ measured at
40 degrees with the MRS spectrometer (scaled by 100 for clarity purposes), followed by the yields of negative hadrons measured at 12 (scaled up by 10) and 4 degrees respectively. The distributions obtained
at $\eta=2.2 $ and 3.2 have variable bin size indicated by horizontal lines, the data points have been located at the mean value calculated within that particular bin. The distribution obtained at $\eta=1$ has fixed bin width of 200 MeV/c and 
the data points are displayed at the middle of the bin.}
\end{figure}

Experimental methods and analysis techniques used to extract the spectra presented here  are described 
in \cite{BRAHMSdA}.
Figure \ref{fig:spectra} shows the invariant yields obtained from p+p collisions (panel a) and d+Au collisions (panel b).
For each system we studied particle production at 40 degrees with the MRS spectrometer and 12 and 4 degrees with the 
FFS spectrometer. Each distribution was obtained from several magnetic field settings and corrected for the 
spectrometer acceptance, tracking and trigger efficiency. No corrections were applied to the spectra for absorption or weak decays. 
Statistical errors are shown as vertical lines, and an overall systematic error of 15\%  is shown as shaded boxes 
on all 
points.
The data points for the spectra extracted at $\eta$ = 3.2 as well of those from $\eta = 2.2$ have been placed 
at the 
mean $p_{T}$ calculated within each bin. This turns out to be equivalent in magnitude to the 
 unfolding of the momentum resolution. The momentum resolution of the spectrometers at full magnetic field is
$\delta p/p = 0.0077 p $ for the MRS spectrometer and $\delta p/p = 0.0018 p$ for the FS with p given in units of GeV/c.
The p+p spectra have also been corrected for trigger efficiency by $13 \pm 5\% $ to make them minimum 
biased with respect to the total inelastic cross section.  
The proton+proton data sample at forward angles was collected in its majority with the magnets set to favor the 
detection of negative particles. 
The spectra at $\eta = 3.2$ have been fitted with a power law function 
$  \frac{C}{(1+\frac{x}{p_0})^n}$ and the integral of that 
function over $p_{T}^2$ is 
compared for consistency in table \ref{tab:one} with UA5 results 
\cite{UA5} for the p+p system, as well as
our own multiplicity measurement for the d+Au system \cite{BRAHMS_daumult}.

\begin{table}
\caption{\label{tab:one}Fits to power law shapes at $\eta = 3.2$.} 

\begin{indented}
\lineup
\item[]\begin{tabular}{@{}*{7}{l}}
\br                              
$\0\0System$&$\frac{dN}{d\eta}_{fit}\0/\0 \frac{dN}{d\eta}_{meas}$&$p_{0}$&\m$n$&$\chi^2/NDF$\cr 
\mr
\0\0$p + p$ & $1.05\pm 0.06\0 /\0  0.95\pm0.07 $ & $1.18\pm0.16$ & $ 10.9 \pm 0.9$ & 13. / 11\cr
\0\0$d + Au$& $2.23\pm0.09\0 /\0 2.1\pm0.6$  & $1.52\pm0.1$  & $ 12.3 \pm 0.5$ & 102 / 11  \cr

\br
\end{tabular}
\end{indented}
\end{table}

\section{The nuclear modification factor}

The d+Au spectra are compared to normalized p+p distributions measured at the same 
spectrometer setting. This comparison with p+p results is based in the assumption that 
the production of moderately high transverse momentum particles scales with the
number of binary collisions $N_{coll}$ in the initial stages. The so called  nuclear modification 
factor is defined as:
\begin{equation}
R_{dAu} \equiv \frac{1}{N_{coll}} \frac{N_{dAu}(p_{T},\eta)}{N_{pp}(p_{T},\eta)}
\label{equation1}
\end{equation}
where $N_{coll} $ is estimated to be equal to $7.2 \pm 0.3$ for minimum bias collisions.

\begin{figure}[!ht]
\begin{center}
\resizebox{0.8\textwidth}{!}
           {\includegraphics{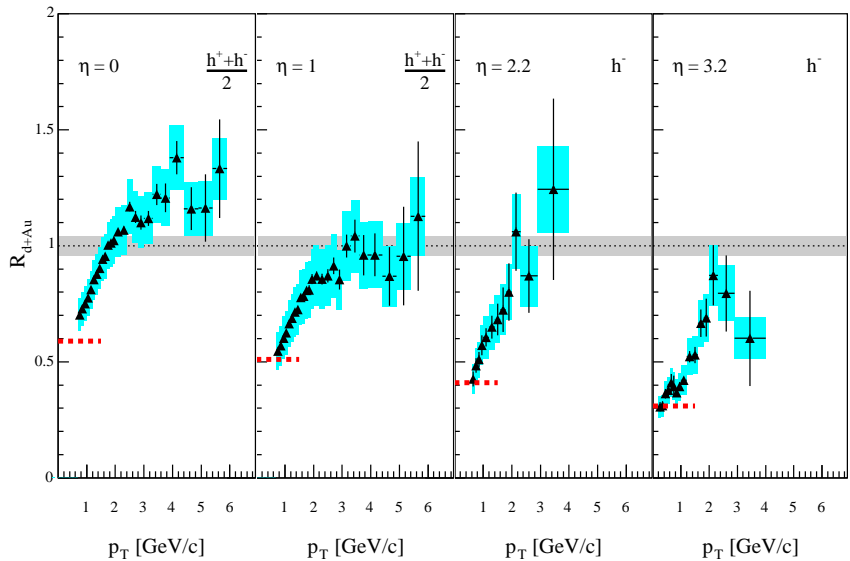}}
\end{center}
\caption{\label{fig:ratio} Nuclear modification factor for charged
  hadrons at pseudorapidities $\eta=0,1.0,2.2,3.2$. Statistical
  errors are shown with error bars. Systematic
  errors are shown with shaded boxes 
  with widths set by the bin sizes.                                          
  The 
  shaded band around
  unity indicates the estimated error on the normalization to $\langle N_{coll} \rangle$. 
  Dashed lines at $p_T<1$ GeV/c show the normalized charged particle 
  density ratio $\frac{1}{\langle
  N_{coll}\rangle}\frac{dN/d\eta(d+Au)}{dN/d\eta(pp)}$.}
\end{figure}

Figure \ref{fig:ratio} shows the nuclear modification factor defined above for four $\eta$ values. The ratio 
at $\eta=0$ and 1 was calculated by averaging on both positive and negative charges $(h^+ + h^-)/2$, 
and the one at $\eta=2.2$ and 3.2
has been calculated including only $h^-$ particles. The average charge is the preferred quantity because it compensates
the isospin mismatches between the systems being compared. Because we do not have enough data to extract 
the distribution of positive charged particles from p+p, the ratio at $\eta=2.2$ and 3.2 was calculated with
spectra of negative charged particles. We can use the model PYTHIA to state that if we were to make the ratio
with the average charge, the ratios would be smaller than those reported in this work.
At mid-rapidity ($\eta = 0$), the nuclear modification factor exceeds 1 for
transverse momenta with values greater than 2 GeV/c in similar way as the measurements
performed by Cronin  at lower energies \cite{CroninEXP}. Such enhancement has been related
to multiple scattering at the partonic level \cite{CroninTheo}.
 
It is seen that a shift of one unit of rapidity is enough to make the Cronin type enhancement disappear, and further 
increases in 
$\eta$ decrease even further the scaled yield of charged particles produced in d+Au collisions.

In all four panels, the statistical errors, shown as error bars (vertical lines), are dominant specially in our most
forward measurements. Systematic errors are shown as shaded rectangles. An additional systematic error is introduced 
in the calculation of the number of collisions $N_{coll}$ that scales the d+Au yields to a nucleon-nucleon 
system.  That error is
shown as a $15\%$ band at $R_{dAu}=1$.  

We see a one to one correspondence between the $R_{dAu}$ values at low $p_{T}$  and the ratio 
 $\frac{1}{\langle
N_{coll}\rangle}\frac{dN/d\eta(d+Au)}{dN/d\eta(pp)}$ as demonstrated in Fig. \ref{fig:ratio} where that ratio
is shown as dashed lines at $p_T <1$. 

\section{Centrality dependence}

The relation between particle yields and the centrality of the d+Au collision has also 
been explored using the fact that
the multiplicity of charged particles in a reasonably wide $\eta$  range ($\mid \eta \mid \leq 2.2$) correlates well to 
the impact parameter of the collision. Three samples of different centralities 
were defined according to
the multiplicity of each event, and scaled histograms in transverse momentum were filled:
 $N_{central}(p_{T}) \equiv \frac{1}{N_{coll}}N_{0-20\%}(p_{T}) $ for  events with multiplicities ranging from 0
 to 20\%.
 $N_{semi-central}(p_{T}) \equiv \frac{1}{N_{coll}}N_{30-50\%}(p_{T}) $ for semi-central 
events with multiplicities ranging from 30 to 50\%, and finally, 
 $N_{periph}(p_{T}) \equiv \frac{1}{N_{coll}}N_{60-80\%}(p_{T}) $ for peripheral events
with multiplicities ranging from 60 to 80\% with $N_{coll}$ values listed in table \ref{tab:two}.

With these histograms, two ratios were constructed: 
$R^{central}_{CP} = \frac{N_{central}(p_{T})} {N_{periph}(p_{T})}$ and 
$R^{semi-central}_{CP} = \frac{N_{semi-central}(p_{T})} {N_{periph}(p_{T})}$.

\begin{figure}[!ht]
\begin{center}
\resizebox{0.8\textwidth}{!}
           {\includegraphics{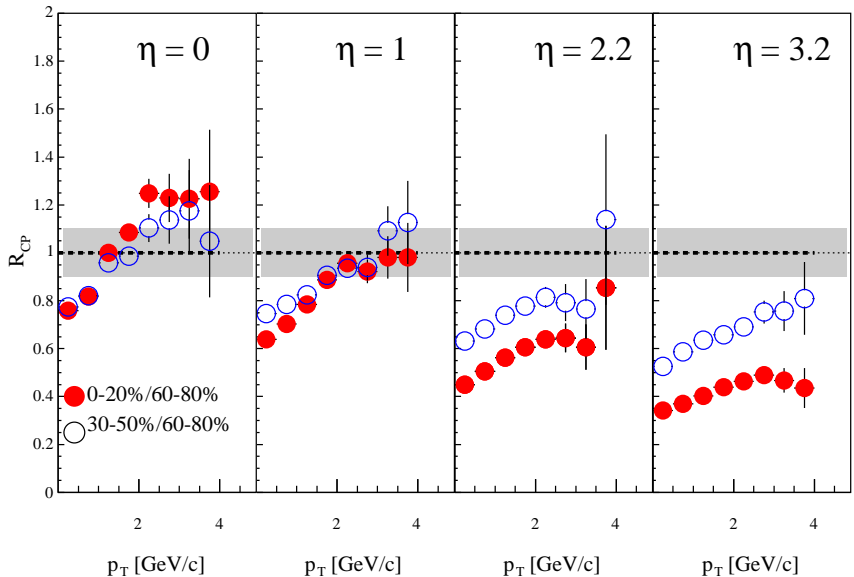}}
\end{center}
\caption{\label{fig:centrality} Central (full points) and
    semi-central (open points) $R_{cp}$ ratios (see text for details)
    at pseudorapidities $\eta=0,1.0,2.2,3.2$. Systematic errors ($\sim5\%$) 
    are smaller than the symbols. The ratios at all pseudorapidities are calculated for the average charge $\frac{h^+ + h^-}{2}$.}
\end{figure}

Because these ratios are constructed with events from the same data run, many corrections cancel out. 
The only correction 
that was applied to these ratios is related to trigger inefficiencies that become important in peripheral 
events. The 
dominant systematic error in these ratios stems from the determination of the average number of binary 
collisions in
each centrality data sample. This error is shown as a shaded band around 1 in Fig. \ref{fig:centrality}.

\begin{table}
\caption{\label{tab:two}$N_{part}$ and $N_{coll}$ values extracted from HIJING calculations for d+Au collisions.} 

\begin{indented}
\lineup
\item[]\begin{tabular}{@{}*{7}{l}}
\br                              
$\0\0Centrality$&$N_{part}(Au)$&$N_{part}(d)$&\m$N_{coll}$\cr 
\mr
\0\0$Central\0 0-20\%$& 12.5  &1.96&$ 13.6 \pm 0.3$\cr
\0\0$Semi-central\0 30-50\%$& 7.36 & 1.79 &$ 7.9 \pm 0.4$&   \cr 
\0\0$Peripheral\0 60-80\%$ & 3.16 & 1.39 &$ 3.3 \pm 0.4$    \cr 

\br
\end{tabular}
\end{indented}
\end{table}

The four panels of figure \ref{fig:centrality} show the central $R^{central}_{CP}$ (filled symbols) and semi-central $R^{s
emi-central}_{CP}$
(open symbols) ratios for the for $\eta$ settings. Starting on the left panel corresponding to $\eta=0$, the 
central events yields are 
systematically higher than those of the semi-central events. On the second panel corresponding to events measured at 
$\eta=1$, the scaled yields of both samples are almost equal. At $\eta=2$ the trend has already reversed, this time the semi-central
events have higher yields than the central events, and finally, at $\eta=3$, the yields of central events are 
$\sim 60\%$ lower than the semi-central events for all values of transverse momenta.

\section{Discussion}

The first results from the 2003 RHIC run showed that Cronin like enhancements are still prominent around mid-rapidity at RHIC 
energies and that jet suppression detected in Au-Au systems \cite{RHIC-AuAu-Supp} is not present in d+Au 
\cite{RHICdA,BRAHMSdA}. It was then concluded that jet suppression must happen in a medium formed after the
interaction of the Au ions, and that medium is partonic and highly opaque.
 
In recent months, several groups reported theoretical discussions as well as numerical calculations
within the context of the CGC with predictions about trends in the physics of deuteron-gold collisions at RHIC. 
Several authors showed that the color glass condensate exhibits a Cronin like enhancement
at mid-rapidity \cite{JamalFirst,Jamal,Dumitru,Kovchegov} while others argued for a gradual disappearance of 
such an 
enhancement as the energy of the collision or the rapidity of the measured particles is 
increased \cite{KKT} or offered numerical calculations based on similar theoretical arguments
 \cite{Wiedemann}. All these theoretical works at higher rapidities introduce a modification of the gluon
densities related to gluon emission probabilities that grow as the rapidity gap between the probe
and the target increases, within a formalism called `quantum evolution'.
All these works predicted behaviors of the nuclear modification factor and its centrality dependence that are
consistent with the measurements being reported in this paper. 

However, other groups based their 
predictions and the description of the data with a two component model that includes a parametrization of 
perturbative QCD and string
breaking as a mechanism to account for coherent soft particle production, the so called HIJING model 
\cite{HIJING}. Multiple scattering at the parton level has been added to this model with different 
degrees of sophistication and it describes well the d+Au results at mid-rapidity as well as the multiplicity 
densities that are dominated by soft processes. These models produce Cronin type enhancements at higher 
rapidities, and the magnitude of that enhancement  grows with centrality. \cite{Vitev}, \cite{XWang}. Such 
behavior is not present in the data presented here.

\begin{figure}[!ht]
\begin{center}
\resizebox{0.8\textwidth}{!}
           {\includegraphics{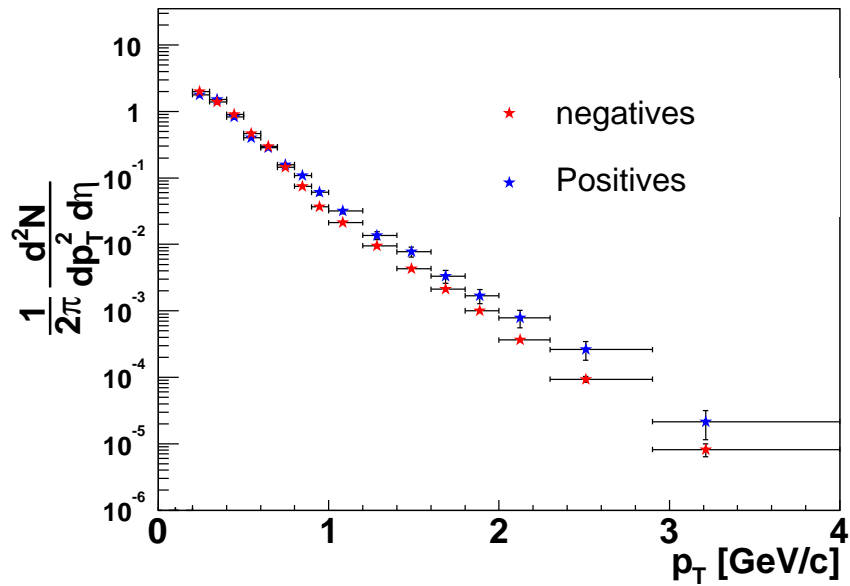}}
\end{center}
\caption{\label{fig:positivedA}Spectra of positive and negative charged particles at $\eta = 3.2$}
\end{figure}

More recently Miklos Gyulassy pointed the higher yield of positive charged particles at $\eta=3.2$ shown
in Fig \ref{fig:positivedA} as
 `clear evidence of valence quark fragmentation' \cite{RIKEN} and the fact that the HIJING model provides a 
good 
representation of the data  with a mechanism of string breaking introduced to explain low energy p-A 
measurements
\cite{Brodsky}. The authors of that paper offer an explanation for the triangular shape of the ratio 
$\frac{dN/d\eta(p+A)}{dN/d\eta(pp)}$ that is also present in these measurements. 

In summary, BRAHMS has studied the production of charged particles
and compared their yields to those produced in proton-proton collisions scaled 
by the number of binary collisions and finds a clear evolution as  
the $\eta$ of the particles changes from mid-rapidity where a Cronin type 
enhancement is seen above transverse momentum of 2 GeV/c, to a gradual 
suppression as the values of $\eta$ change to 1, 2.2 and 3.2.
We have also found that the suppression increases with the centrality of the
collision.

\section{Acknowledgments}

This work was supported by 
the Office of Nuclear Physics of the U.S. Department of Energy, 
the Danish Natural Science Research Council, 
the Research Council of Norway, 
the Polish State Committee for Scientific Research (KBN) 
and the Romanian Ministry of Research.

\end{document}